\documentclass[letterpaper]{article}
\usepackage{aaai20}
\usepackage{times}
\usepackage{helvet}
\usepackage{courier}
\usepackage{graphicx}
\usepackage{color}
\usepackage{caption}
\usepackage{subcaption}
\usepackage{enumitem}
\usepackage{nicefrac}
\usepackage{soul}
\begin{document}

\title{Soliciting Human-in-the-Loop User Feedback for Interactive Machine Learning Reduces User Trust and Impressions of Model Accuracy}

\author{Donald R. Honeycutt, Mahsan Nourani, Eric D. Ragan\\
University of Florida, Gainesville, Florida\\
dhoneycutt@ufl.edu, mahsannourani@ufl.edu, eragan@ufl.edu\\
}

\maketitle
\begin{abstract}
Mixed-initiative systems allow users to interactively provide feedback to potentially improve system performance.
Human feedback can correct model errors and update model parameters to dynamically adapt to changing data.
Additionally, many users desire the ability to have a greater level of control and fix perceived flaws in systems they rely on.
However, how the ability to provide feedback to autonomous systems influences user trust is a largely unexplored area of research.
Our research investigates how the act of providing feedback can affect user understanding of an intelligent system and its accuracy.
We present a controlled experiment using a simulated object detection system with image data to study the effects of interactive feedback collection on user impressions.
The results show that providing human-in-the-loop feedback lowered both participants' trust in the system and their perception of system accuracy, regardless of whether the system accuracy improved in response to their feedback.
These results highlight the importance of considering the effects of allowing end-user feedback on user trust when designing intelligent systems.
\end{abstract}

\section{Introduction}
\label{sec:intro}
%Getting human input in ML models is good
Bringing human feedback into the development of machine learning models has many benefits.
At its simplest, human feedback allows a model to incorporate new annotations for unlabeled data to increase performance by improving the training set.
A common method for introducing human feedback is active learning, where the selection of data to obtain labels for is left to the model~\cite{cohn1996active}.
Alternatively, a more human-centered approach has the labeler choose which instances to be labeled, relying on human intuition to decide what feedback would be most relevant to improve the model based on observations of its performance~\cite{tong2001support}.
Developers can also allow for further involvement by giving the human participant feature-level control over model parameters, such as allowing direct modification of the feature space and its associated weights~\cite{cho2019explanatory} or prioritizing decision rules used by the model~\cite{yang2019study}.

%End users of systems can be involved in the feedback loop
Frequently, the human-in-the-loop approaches either involve system developers for development and debugging~\cite{vathoopan2016human} or independent workers on crowd-sourcing platforms~\cite{li2017human}.
By taking advantage of end-users' periodical feedback upon noticing errors, these models can stay updated in the presence of shifting data or changing goals.~\cite{geng2009incremental,yamauchi2009optimal,elwell2011incremental}.
Systems can also update over time by implicitly capturing user behaviors, which is a technique commonly used in recommender systems~\cite{shivaswamy2012online,middleton2003capturing}.
While this feedback is not provided explicitly, users can still observe the system directly reacting in response to their actions, decisions, and feedback.
Furthermore, end users of intelligent systems may want the ability to correct observed model errors.
When engaged with the outcomes of a system, many users desire the ability to influence those outcomes by providing feedback beyond simple error correction~\cite{stumpf2008integrating}.

%T
While human-in-the-loop systems can have improved model accuracy and provide users control over the systems they rely on, there may also be unexplored consequences to allowing end users to provide feedback.
For instance, Van den Bos et al.~\shortcite{van1996consistency} observed that when interacting with human teams, the ability to provide feedback has been observed to have a positive effect on the perceived fairness of team decisions.
In their study, users who felt their feedback was considered reported higher levels of trust in the decision-making process and were more committed that the correct decision was made.
They also observed the inverse effect, with a decrease in trust in the team if feedback was provided but ignored~\cite{korsgaard1995building}.
Since providing feedback to an automated decision-making system is similar to providing feedback to a human-based decision making system, it is possible that similar effects could be observed in human-in-the-loop systems.

If providing feedback does affect user trust, it could lead to people misusing the systems they provide feedback to.
When experiencing a higher level of trust than is appropriate based on the system performance, users may over rely on the system.
On the other hand, having a lower level of trust could result in not using the system at all~\cite{lee2004trust,parasuraman1997humans}.
Therefore, it is important to understand how providing feedback to an intelligent system affects trust so that it can be accounted for when designing human-in-the-loop systems.

%What we did
In this paper, we examine how users perceive system accuracy over time and how their trust in the system changes based on the presence of interactive feedback.
We used a simulated object-detection system that allowed users to provide interactive feedback to correct system errors by adjusting image regions for detected objects.
Additionally, to explore possible implications of how the system responds to given feedback, our experiment also controlled different types of change in system accuracy over time.
The results indicate that by providing human-in-the-loop feedback, user trust and perception of accuracy can be negatively affected---regardless of whether the system improves after receiving feedback.

\section{Related Work}
\label{sec:related}
In this section, we consider prior work from the perspectives of human-in-the-loop machine learning and trust in artificial intelligence.

%Active learning/human-in-the-loop/relevance feedback (usefulness of human feedback in ML models)
\subsection{Human-in-the-Loop Machine Learning}

While machine learning can be used to train models based purely on data without direct human guidance, there are many scenarios where incorporating human feedback is beneficial.
In many cases, this feedback is simply having a human annotate new data to be incorporated into the model.
\textit{Relevance feedback} is a human-in-the-loop method where a human reviews the pool of unlabeled data alongside the current model's predictions on that data, choosing when to provide new labels to the system based on their own intuition~\cite{tong2001support}.
Another approach that can be taken in domains where human intuition may not result in optimal selections of what data to label is to choose instances to add to the training set by objective metrics based on the model.
Active learning selects relevant instances to show to a human, referred to as an \textit{oracle}, based on which unlabeled data are most likely to represent information missing in the current version of the model~\cite{cohn1996active}.
While theoretical active learning research treats the oracle as merely being a way to obtain the true labels for selected data, in practice, active learning models need to account for the fact that the oracle is a human and therefore not infallible~\cite{settles2011theories}.

%Interactive learning (types of human feedback beyond simple annotation)
However, human input is not limited to merely providing new labels to data.
Explanatory interactive learning has the oracle not only provide the appropriate label for the data point but also provides an explanation of the current model prediction and asks the oracle to correct the reasoning in the explanations~\cite{teso2019explanatory}.
This helps avoid situations where the model has a flaw that happens to result in the correct prediction by chance.
Another form of advanced human feedback is to show the oracle model parameters, such as features and their weights~\cite{cho2019explanatory} or rules used to make decisions within the model~\cite{yang2019study}, and allows for direct modification of those parameters.
Being able to control model parameters in this way has been found to be useful for debugging models~\cite{kulesza2010explanatory}.
While this higher level of control over the model may not be desirable in all applications, Holzinger et al.~\shortcite{holzinger2016towards} showed that human-machine teaming can sometimes result in a closer to optimal model than machine learning alone.

%Incremental learning/concept drift/coactive learning (benefits of end user feedback/continuously updating the model after its in use, reason that trust of people providing feedback is relevant)
While the person providing feedback is not necessarily the end user for many human-in-the-loop systems, there are advantages to bringing end users into the loop.
Stumpf et al.~\shortcite{stumpf2008integrating} found that users of intelligent systems largely want to provide feedback to systems they are using, particularly when it gives them a feeling of being able to control some aspect of the model.
Similarly, people are more likely to use an imperfect intelligent system when they have the ability to correct its errors \cite{dietvorst2018overcoming}.
Additionally, end users may notice when an already deployed system begins to falter.
Even if a model was very accurate at the initial time of training, the training data may become less representative of the actual population of data as trends shift over time.
This is a phenomenon known as concept drift~\cite{vzliobaite2010learning}.
A human-in-the-loop approach to dealing with this problem is known as \textit{incremental learning}, where the model periodically obtains labels as they become available to update the model while it is in use~\cite{geng2009incremental}.
These techniques have been shown to effectively address the problem of concept drift in machine learning systems~\cite{yamauchi2009optimal,elwell2011incremental}.

%Transition into trust and how providing feedback could affect how users perceive systems (effects of usage of input on trust in the psychology of collaborative decision making)
Providing input has been shown to affect trust and perception of fairness in the field of psychology.
In decision-making teams, people were observed to place more trust in a team-leader who actively considered their input, and they were also more confident that the correct decision was made after the fact~\cite{korsgaard1995building}.
A similar effect was observed in procedural decision making systems, with people having a higher level of trust and perception of fairness in a decision-making system that they were able to give input to~\cite{van1996consistency}.
An interesting result from both of these studies was that providing feedback had a negative effect on trust if the feedback was ignored.
Since feedback affects interpersonal trust by improving trust if feedback is considered and decreasing trust if it is ignored, a similar effect may be observed in human-computer interactions.

\begin{figure*}[ht]
    \centering
    \includegraphics[width=0.99\textwidth]{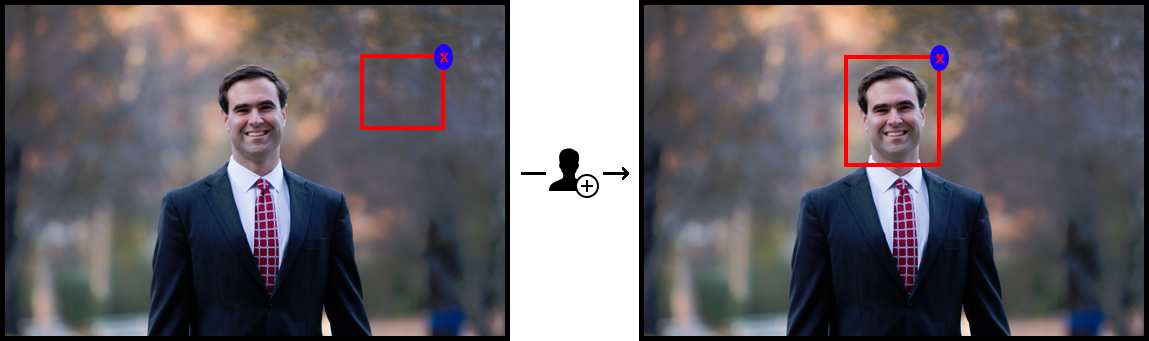}
    \caption{In the \textit{with interaction} conditions, participants could delete existing bounding boxes or click-and-drag to create new ones.
    In this example, the left image shows a system error, and the right shows a version after interactive correction. \footnotemark
    }
    \label{fig:interface}
\end{figure*}

\subsection{Trust in Artificial Intelligence}

User trust in artificial intelligence systems has been studied for many years and is of value since it is directly associated with usage and reliance~\cite{parasuraman1997humans,siau2018building}.
As a result, users need to place an appropriate amount of trust in a system based on its performance in different contexts.
Reliance and trust in automated systems are not binary (i.e., to trust or not) and are generally more complex~\cite{lewicki1998trust,lee2004trust}.
Desired behavior is for a user to examine a system's outputs and decide whether to rely on the system based on the accuracy of results~\cite{hoffman2013trust}.
This behavior has been observed to be more prevalent among users who are domain experts than novice users~\cite{nourani2020Role}.
Sometimes, however, users might trust a system completely without checking the outcomes, i.e., over-reliance or \textit{automation bias}~\cite{goddard2012automation}.
This situation can be caused by a user's lack of confidence or when the system seems more intelligent than they are  based on their initial preconceptions~\cite{lee2004trust,hoffman2013trust,nourani2020investigating}.
In contrasting scenarios, \textit{mistrust}~\cite{parasuraman1997humans} and \textit{distrust}~\cite{lee2004trust} can cause users to rely more on themselves or under-rely on the system.
Both of these situations can be dangerous, especially for systems with critical tasks where decisions can be fatal.
For example, wrong decisions in criminal forecast systems can wrongfully convict an innocent person~\cite{berk2015machine}.

To raise users' trust and provide more information to aid them in their decision-making process, researchers have explored the use of explainability in artificial intelligence systems~\cite{ribeiro2016should}.
Studies of human-in-the-loop paradigms have shown explainability can help users understand and build trust in the algorithms in order to provide proper feedback and annotations~\cite{ghai2020explainable,teso2018should}.

Researchers use different methods to measure trust and reliability in machine learning and artificial intelligence systems.
For example, some researchers utilize user's agreement with the system outputs as a measure of reliance and trust;
specifically, identifying when the user agrees with the system outputs that are not correct~\cite{nourani2020don}. 
Yu et al.~\shortcite{yu2019trust} propose a  \textit{reliance rate} based on the number of times the users agreed with the system answers out of all their decisions.
In recent work, Yin et al.~\shortcite{yin2019understanding} found that trust is directly affected by user's estimations of the system's accuracy, where underestimation of accuracy can cause mistrust in the system, and vice versa.
As a result, a user's estimated or observed accuracy can be used as an indirect measurement for user trust, and we use these methods in the study reported in this paper.
\footnotetext{Image from ``Josh McMahon Portraits - 2517'' by John Trainor (used under CC BY 2.0) with annotations added by the authors. License available at https://creativecommons.org/licenses/by/2.0/}

\section{Method}
\label{sec:method}
In this section, we discuss our research objectives based around understanding the differences in user trust among users of human-in-the-loop systems and non-interactive systems.
We also present details of our experimental design and study procedure.

\subsection{Research Objectives}
With the goal of understanding the effects of providing human-in-the-loop feedback on user perception of artificially intelligent systems, we identified the following research questions:
\begin{enumerate}[align=left]
    \item [\textbf{RQ1:}] Does user trust in an intelligent system change if the user provides feedback to the system? 
    \item [\textbf{RQ2:}] Does providing feedback to an intelligent system affect user ability to detect changes in system accuracy over time?
\end{enumerate}

To address these research questions, we designed a controlled experiment using a simulated image classification system both with and without feedback.
With these different systems, we hypothesized that the effects of participant trust due to interaction presence would be based on system response to their feedback, similar to the effects observed in human-based decision making systems~\cite{korsgaard1995building,van1996consistency}.
If the system reacted positively, improving as the participant provided feedback, we expected that participants would feel more invested in the system and as a result, they would trust the system more.
However, if the system did not honor the user's feedback and did not improve after taking participant feedback, we expected that they would become negatively biased against the system.

% Experimental Design
\subsection{Experimental Design}

For our study, we provided participants with a series of images with classifications from a simulated model.
To avoid confusion of whether a system prediction was correct or not, we chose to focus on a domain which required no prior experience, which led us to use detection of human faces as our classification goal.
With the goal of increasing participant engagement in the feedback process, we wanted to use a system with more intricate outputs than binary classification alone. 
Therefore, we decided to simulate a system that detected the location of human faces rather than just their presence, placing bounding boxes over each face in the image.
Classifications were hand-crafted, not actually from an artificially intelligent model as participants were told.
The task consisted of reviewing three rounds of images, with 30 images in each round.
The images used in our simulated model were taken from the Open Images dataset~\cite{OpenImages,OpenImages2} with our own manually generated annotations.
Each round of 30 images contained 20 pictures of people, with the remaining 10 images containing things such as animals or empty scenery.
For images where we chose to simulate system errors, we used a roughly equivalent mix of false positives (bounding boxes placed on objects that were not human faces) and false negatives (unidentified human faces).

%  Longer description of system
Because our main metrics---perception of model accuracy and user trust---are based on participants' experiences with the system, we decided that each participant should only see one version of the system so as not to be biased by their experience with the previous system versions.
For this reason, we used a 2x3 between-subjects design for the experiment.
The first independent variable in our experiment was \textit{interaction presence} with two levels: \textit{with interaction} and \textit{without interaction}.
Participants in the \textit{with interaction} condition were asked to provide feedback to the model for each image by correcting any errors, or by verifying that the system's classification was correct.
To do this, participants could interact with the system to removing any bounding boxes from an image that did not contain a human face, and they could add new bounding boxes over any unidentified faces in the image. 
Participants in \textit{with interaction} condition were explicitly instructed that their feedback would be used by the model in-between each round of images to update the model's parameters before classifying the next round of images.
To maintain a feeling of realism that the model was actually updating, we added a 45 second pause between each round and told participants that they would need to wait for the model to take their feedback into account and update the predictions for future classifications.

The \textit{without interaction} system removed the ability to interact with the bounding boxes to correct erroneous instances.
In both conditions, to ensure participant engagement in this condition, we asked participants to respond whether the model's classification was correct or incorrect.
Unlike the participants who saw the interactive system, participants in the \textit{without interaction} condition were told that their responses would be sent to the researchers after the completion of the final round of images, with no indication that their responses would be used by the model in any way.

\begin{table}[b]
    \resizebox{\columnwidth}{!}{
    \def\arraystretch{1.2}
    \begin{tabular}{|l|c|c|c|}
    \hline 
         & \textbf{Round 1} & \textbf{Round 2} & \textbf{Round 3} \\ \hline
         \textit{Increasing} accuracy & 70\% & 80\% & 90\% \\ \hline
         \textit{Constant} accuracy & 80\% & 80\% & 80\% \\ \hline
         \textit{Decreasing} accuracy & 90\% & 80\% & 70\% \\ \hline
    \end{tabular} }
    \caption{System accuracy by round.}
    \label{tab:qual}
\end{table}

Our second independent variable was \textit{change in accuracy}, which corresponded to the simulated accuracy of the system in each round of images with levels as shown in Table \ref{tab:qual}:

While the change in accuracy factor influenced the distribution of errors over sections of the study, it is important to note that the \textit{total} number of system errors observed by participants across the entire study was the same in all conditions---a total of 18 of 90  images were shown as classified incorrectly regardless of condition.
The only difference among these conditions was when those errors were shown.

\begin{figure}[tb]
    \centering
    \includegraphics[width=0.95\columnwidth]{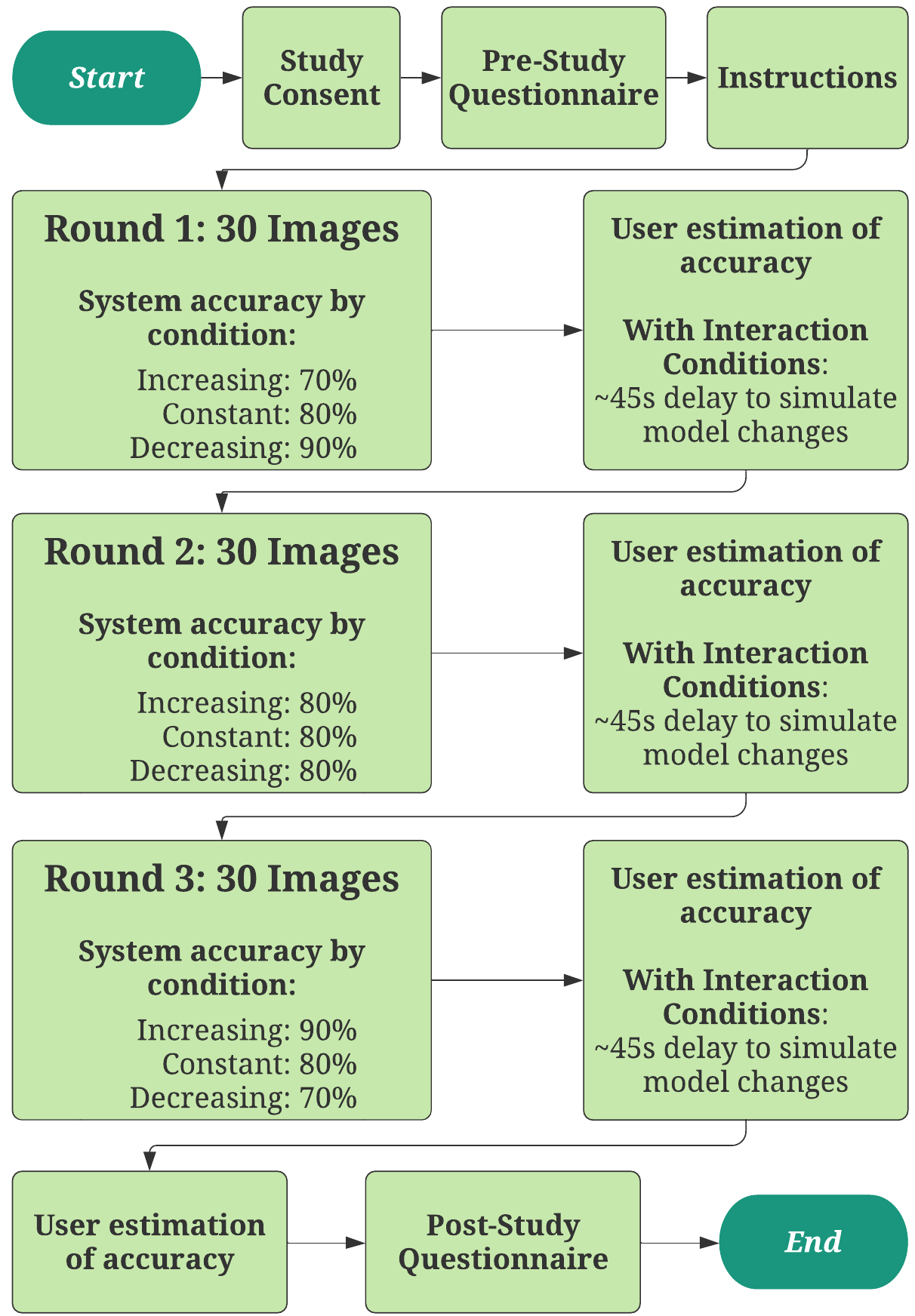}
    \caption{Study procedure overview.}
    \label{fig:procedure}
\end{figure}

\begin{figure*}[tb]
    \centering
    \includegraphics[width=\textwidth]{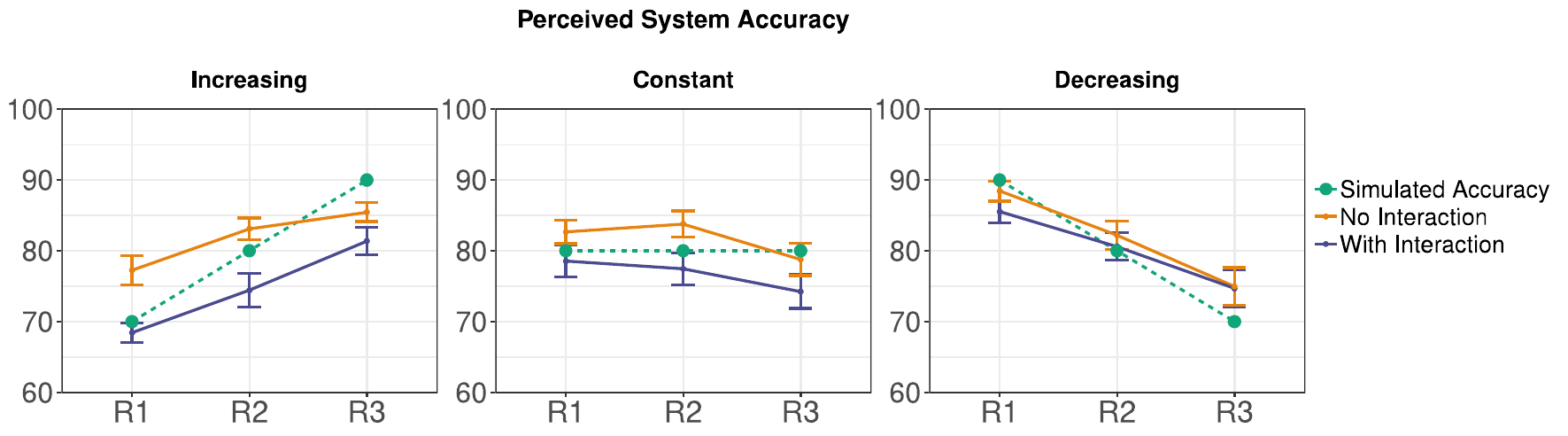}
    \caption{Perceived accuracy across rounds (error bars show standard error). 
    Participants \textit{with interaction} (purple) rated the system as less accurate than those with \textit{no interaction} (yellow).}
    \label{fig:acc}
\end{figure*}

\subsection{Procedure}

Participants completed the experiment using an online web application without intervention or live communication with the researchers.
The study began with a pre-study questionnaire, asking basic demographic information including age, gender identification, and educational background.
Additionally, we asked participants to self-report their experience with machine learning and artificially intelligent systems to ensure there was no significant difference between the experience of the populations for each condition.
Participants then received instructions on completing the task, including examples of correct and incorrect classifications of images.
To avoid ambiguity as to what constituted a correct classification, we instructed participants to consider any bounding box that contained a portion of a human face to be correct.
Additionally, when designing the outputs of the simulated model we avoided placing any bounding boxes that only partially contained a face.
Participants in conditions with interaction also saw a tutorial on how to edit the bounding boxes to provide feedback to the model which reminded them that their feedback would be used to update the model between rounds.

After finishing the instructions and tutorial, participants moved on to the main task, which consisted of three rounds of reviewing 30 images with bounding boxes corresponding to the system classifications.
Between each set of images, participants were asked to estimate how accurate the system was during the previous set of images.
Participants in the \textit{with interaction} conditions were required to wait for an added time delay before being able to continue to the next round (simulating the time required for the system to update based on participant feedback). 
A notification about the reason for this delay was also shown to remind participants that their feedback was being used dynamically (although the actual system remained static regardless of their feedback).
After all rounds of images were completed, participants filled out a post-study questionnaire to evaluate their level of trust in the system.

\subsection{Participants}

Participants were recruited from Amazon Mechanical Turk with a requirement for participants to have the Masters qualification, an approval rate of greater than $90\%$, and $500$ or more prior tasks completed successfully.
Participants ranged from ages 24--68 and lived in the United States at the time of study completion.
To ensure the quality of participant responses, we measured the percentage of responses for which participants correctly identified whether an image corresponded to a system error or not.
As a quality check, participants were not included in the results if they had less than $75\%$ accuracy for either correct instances or system errors.
Our study had a total of 157 participants, and 4 were removed based on the accuracy criteria.
The remaining 153 participants consisted of 83 males, 69 females, and one non-binary response.
Participants took approximately 14 minutes on average to complete the study.

\section{Results}
\label{sec:results}
In this section, we present the measures of our study and empirical results.
We report statistical test results along with generalized eta squared ($\eta^2_G$) for effect sizes of ANOVA tests and Cohen's d ($d_s$) for effect sizes of post-hoc tests.

\subsection{User-Perceived Model Accuracy}

To examine the effects of providing interactive feedback on user-perceived system accuracy, the participants numerically estimated the system accuracy after each round of image review. 
To account for differences in observed accuracy controlled by the \textit{change in accuracy} factor, we analyzed estimated accuracy as the error of participant responses compared to the actual simulated accuracy.
Results are shown in Figure \ref{fig:acc}.
A three-way mixed-design ANOVA was performed on the error of estimated accuracy, with \textit{change in accuracy} and \textit{interaction presence} as between subjects factors and image set (i.e., first, second, or third round) as a within subjects factor. 
The analysis showed a significant effect of interaction presence on error of estimated accuracy.
Participants who provided system feedback estimated the system as being less accurate than those who did not provide feedback to the model, with $F(1,147)=6.99$, $p<0.01$, $\eta_G^2 = 0.035$.
No significant interaction effects were detected.

Additionally, the ANOVA test found participant error to be significantly different based on round with $F(2,294) = 10.29, p<0.001, \eta_G^2=0.016$, as well as an interaction effect between change in accuracy and round with $F(4,294)=38.19$, $p<0.001$, $\eta_G^2=0.108$.
As the simulated accuracy in each round was different based on the change in accuracy condition, these results are not surprising.

\subsection{Perception of Model Change}

\begin{figure}[tb]
    \centering
    \includegraphics[width=\columnwidth]{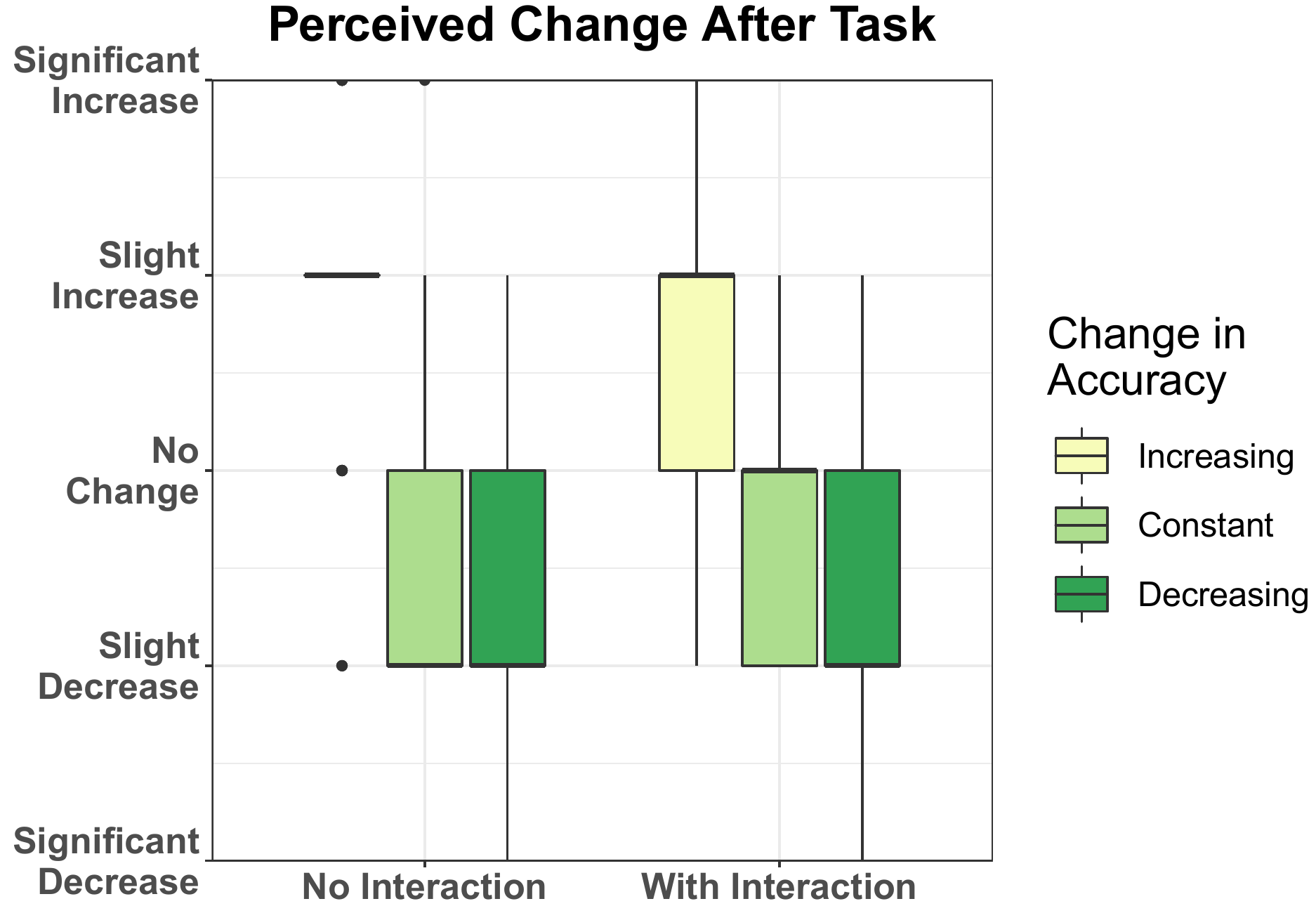}
    \caption{Perceived change in system accuracy from the first round to the last. Participants who saw an increase in accuracy reported a significantly more positive perceived change in accuracy than those who observed constant accuracy, and those who saw constant accuracy reported a significantly more positive change than those who saw decreasing accuracy.}
    \label{fig:changeScale}
\end{figure}

In addition to the reported accuracy during the task, we asked participants to rate how much they thought the system had changed across the different rounds on a five-point Likert scale.
Figure \ref{fig:changeScale} shows the distribution of participant responses to this measure.
We performed an independent two-way factorial ANOVA on participant responses that showed no significance based on interaction presence. 
However, we did observe that change in accuracy was significant with $F(2,147)$, $p<0.001$, $\eta_G^2=0.405$.
A Tukey posthoc test showed that each pair was significantly different.
Participants who saw increasing accuracy rated the system as having changed significantly more positively than both constant accuracy ($p<0.001$, $d_s=1.366$) and decreasing accuracy ($p<0.001$, $d_s=1.946$).
Those who saw constant accuracy thought that the system had a more positive rate of change than participants who observed decreasing accuracy ($p<0.05$, $d_s=0.498$).

\subsection{User Trust}

To measure participants' trust, we asked participants to rate their agreement using a series of scales proposed by Madsen and Gregor that focus on capturing different aspects of human-computer trust~\cite{madsen2000measuring}.
Participants rated each item on a seven-point Likert scale.
Because the scales were developed for intelligent systems that aid in user decision making, we selected the following subset that applied most to our system.
The following three statements were shown to all participants, and the aggregate rating was used as a measure for trust:

\begin{itemize}
    \item The system performs reliably.
    \item The outputs the system produces are as good as that which a highly competent person could produce.
    \item It is easy to follow what the system does.
\end{itemize}

Additionally, as a simple measure of participants' thoughts on the model updating with feedback, the following was only shown to participants in conditions with interaction:

\begin{itemize}
    \item The system correctly uses the information I enter.
\end{itemize}

\begin{figure}[tb]
    \centering
    \includegraphics[width=\columnwidth]{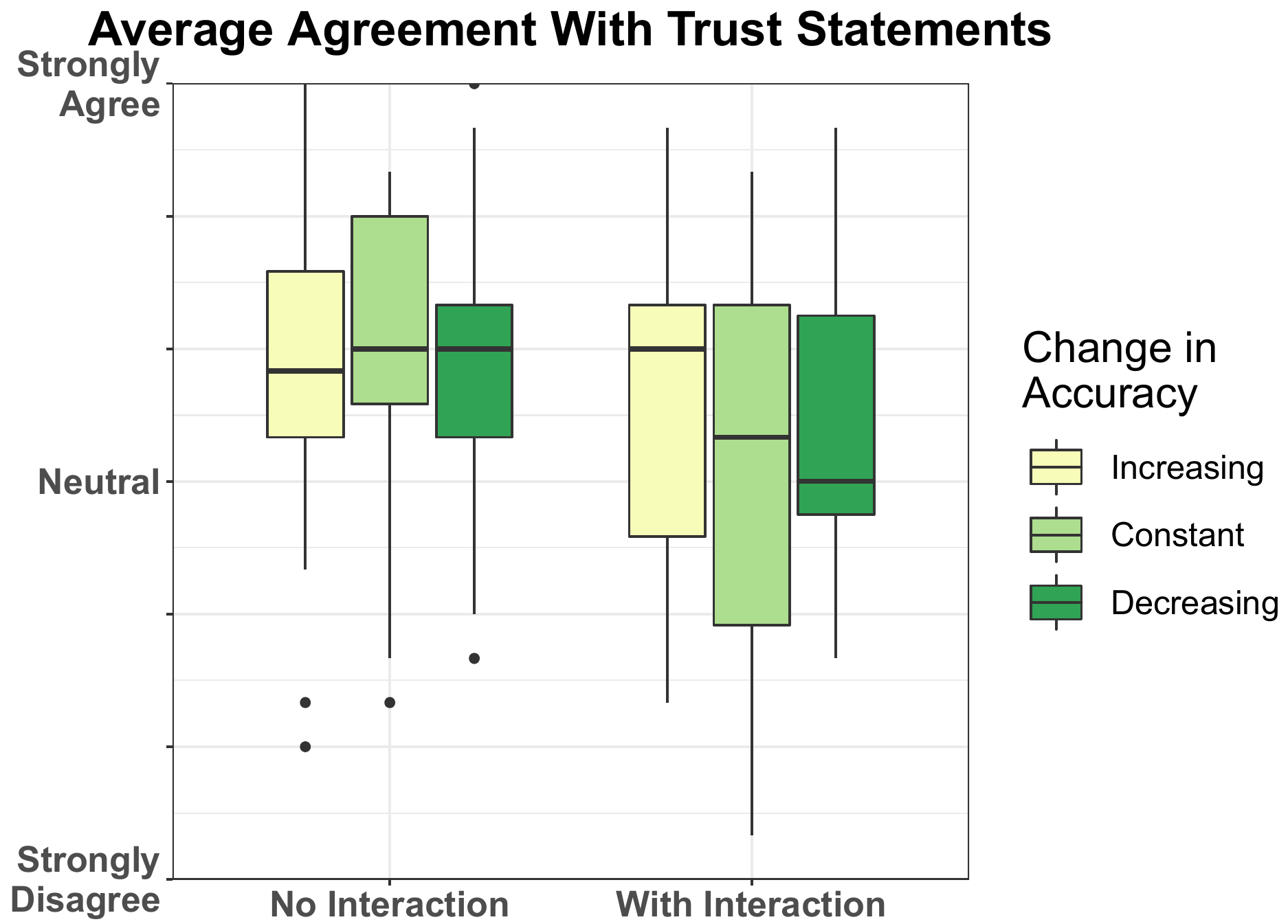}
    \caption{Average agreement with the three trust statements. Participants with interactions had significantly lower trust, regardless of their observation of change in accuracy.}
    \label{fig:trustAvg}
\end{figure}

Aggregated responses for the first three trust items were analyzed with a two-way factorial ANOVA testing the effects of interaction presence and accuracy change.
This test showed that the \textit{with interaction} condition had significantly lower trust than the \textit{without interaction} condition with $F(1,147)=7.61$, $p<0.01$, and $\eta_G^2=0.049$.
The test did not detect a significant effect of \textit{change in accuracy} on participant trust.
The distribution of average participant agreement with the first three trust statements is shown in Figure \ref{fig:trustAvg}.

\begin{figure}[tb]
    \centering
    \includegraphics[width=0.75\columnwidth]{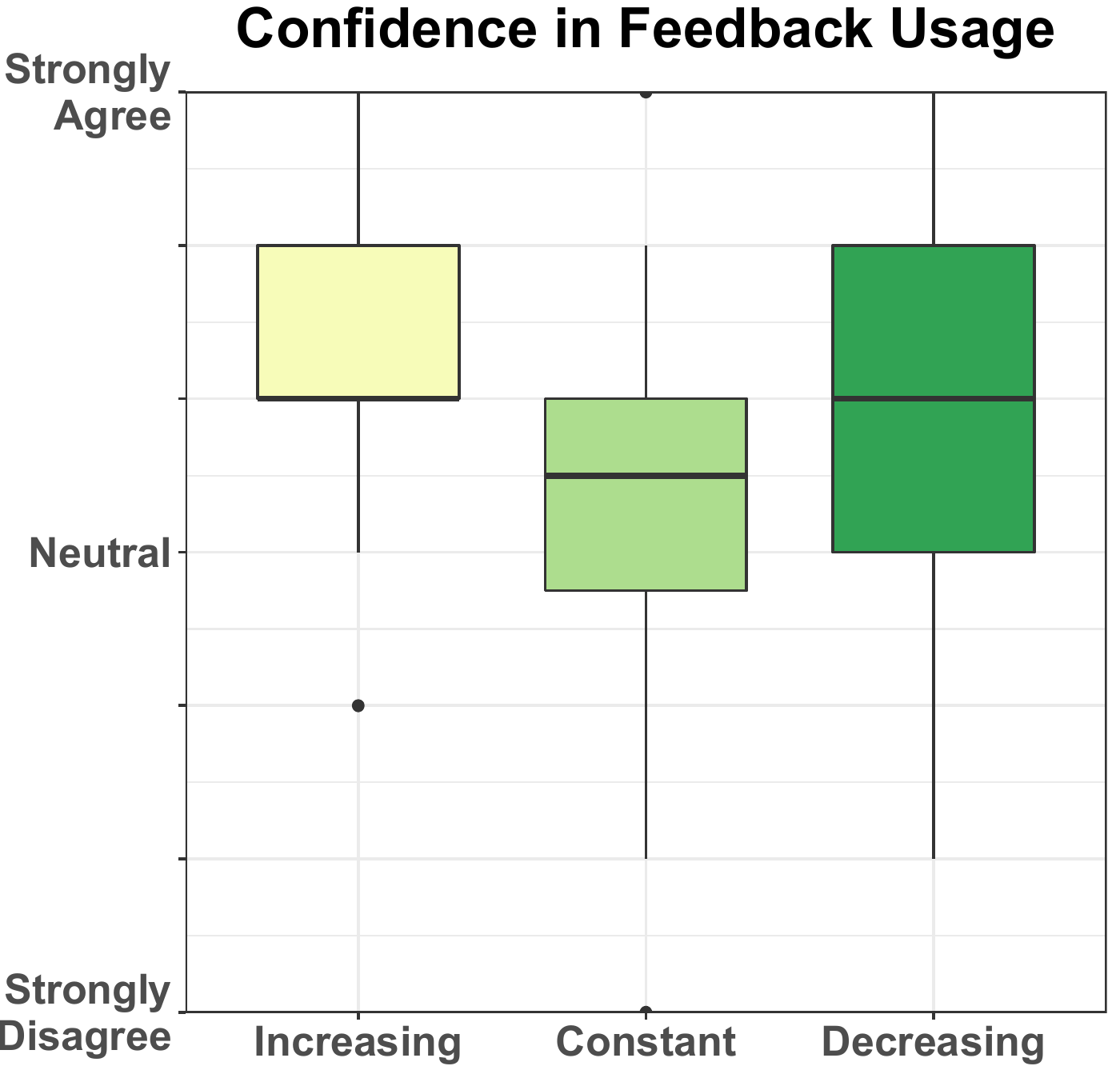}
    \caption{Participant agreement that the system correctly used their feedback. Participants with increased accuracy were significantly more confident in correctness of feedback usage than those with constant accuracy.
    }
    \label{fig:trustUses}
\end{figure}

The results from the fourth statement about agreement that the system correctly used their feedback are shown in  Figure \ref{fig:trustUses}.
Since this measure was only relevant and collected for participants in the \textit{with interaction} conditions, we performed a one-way ANOVA test with \textit{change in accuracy} as the only factor.
The test revealed a significant effect with $F(2,75)=3.263$, $p<0.05$, $\eta_G^2=0.080$.
From a Tukey posthoc test, participants who observed an increase in system accuracy had a higher level of agreement that the system was updating correctly than those with constant accuracy, with $p<0.05$, $d_s=0.725$.
Thus, participants believed their feedback was being used when they corrected the image detection and observed increased accuracy over trial rounds.
We did not observe any significant effect between participants who saw a decrease in system accuracy and participants in either of the other conditions.

\section{Discussion}
\label{sec:discuss}
%Introduction to discussion
This section discusses the results of the experiment in the context of our research questions and hypotheses.
We also consider limitations of our experiment and opportunities for further work on this subject.

\subsection{Interpretation of Results}

Our goal for this study was to explore the effects that providing feedback to an automated system has on both user trust and perception of system accuracy.
In our experiment, we controlled for both presence of interaction and change in system accuracy.
We expected that participants who saw a positive response to their input---an increase in accuracy---would experience an increase in trust and perceived accuracy compared to participants who did not provide feedback.
For participants who did not observe a positive response---constant accuracy or a decrease in accuracy---we expected the opposite.
However, while our analysis did detect a significant effect for presence of interaction on both perceived accuracy and trust, the effect did not depend on the observed change in accuracy as expected.
Rather, participants who provided feedback to the system perceived the system as less accurate and had less trust compared to those who did not provide feedback, regardless of the observed change in accuracy.

This leads us to believe that the observed decrease in user trust may be due to an increase in the salience of system errors.
By correcting the mistakes made by the system, those who provided feedback spent more effort on system errors than those using the non-interactive system.
Since disagreement resulted in participants taking action while agreement did not (and was therefore reviewed faster), memory of disagreements may have been reinforced in the participants' minds.
That is, participants may have remembered the system's mistakes more strongly than the instances where they agreed.
Participants may have also considered the act of providing feedback as an inconvenience, as correcting the system required more time and effort than simply observing whether the system was right or wrong.
The increased memorability of system errors might help explain why participants who used the interactive system trusted it less than those who used the non-interactive version.
This interpretation is also supported by the results of the responses to the trust questionnaire, as trust can be strongly influenced by observed accuracy~\cite{yin2019understanding}.

We also expected a positive change in system accuracy would correspond to higher confidence in feedback usage.
We observed this effect between the increasing and constant conditions, where participants who observed an increase in system accuracy were more confident that their feedback was used correctly than those who observed a constant level of accuracy.
However, no significant difference was observed when comparing participants from increasing and decreasing conditions.
One possible interpretation of this result is that providing more feedback also increases perception of feedback usage.
Participants in the decreasing condition provided the most feedback in the last trial of the task.
As a result of the trial being the closest to the final questionnaire, these participants may have remembered giving the system more feedback than participants in the other conditions, negating the effect that seeing a decrease in accuracy could have.

\subsection{Implication for Human-in-the-Loop Systems}
The findings of this study highlight the importance of thoughtful design of feedback systems.
While these systems can benefit from user feedback to improve model performance, they can also negatively affect trust and perception of accuracy for their users.
This distrust of the system can lead to humans self-relying for critical decisions, which in many cases will result in a higher rate of human error.
For instance, Parasuraman et al.~\shortcite{parasuraman1997humans} report a case where train operators disabled automated alarms due to distrust, which resulted in a significant increase in accidents.
Therefore, designers of human-in-the-loop systems should consider ways to avoid biasing the trust of their users who give feedback.

One potential way to reduce bias due to over-emphasizing attention to errors could be to capture user feedback implicitly.
This technique is commonly found in recommender systems, where the system updates its recommendations based on prior user behavior to improve the relevance of their results.
For example, users perceive search engine results as more relevant if the search engine prioritizes results that were clicked on by prior users~\cite{shivaswamy2012online}.
Similarly, Middleton et al.~\shortcite{middleton2003capturing} designed a system that recommended research papers based on identifying similar users and found that it was effective at making relevant recommendations.
However, they also found that their recommendations were significantly improved when users provided explicit feedback of the topics they were interested in, suggesting that implicit feedback may not always be an adequate replacement for explicit feedback.
While implicit feedback might address the issue of feedback making system errors more salient, it is important to note that systems which operate with implicit updates may open new possibilities for other forms of bias.
By focusing on recommendations that are similar to those which were previously used, the scope of system recommendations for each user can become increasingly narrow over time \cite{de2015investigation}.
Furthermore, if the user of such systems is unaware that this is happening, they can become biased by only being shown content that matches their current beliefs \cite{knijnenburg2016recommender}.

Another approach developers have taken to increase user trust and facilitate an accurate understanding of system accuracy is to introduce explanations of system behavior.
Adding system explanations to interactive systems results in a more appropriate level of user trust, increasing trust when the system is accurate and lowering trust for inaccurate systems~\cite{teso2018should}.
Ribeiro et al.~\shortcite{ribeiro2016should} also showed that by explaining the features used by a classifier in making a prediction, users identified the system accuracy more precisely.
They also found that model accuracy was improved by having users provide feedback to the explanations by removing features that they deemed unimportant.
This suggests that explanations can not only help offset distrust caused by providing  in human-in-the-loop systems, but that explanations may provide further interaction modes and improve the quality of feedback given.

Finally, it may be beneficial to directly make users aware of their potential biases.
Wall et al.~\shortcite{wall2017warning} proposed a series of metrics to detect potential biases by focusing on patterns in what data has been observed by the user.
This can be used to help users identify potential biases towards their trust or understanding of a system based on how much attention they have given to different types of system outputs.
For example, human-in-the-loop systems could be accompanied by visual analytics tools that notify users when they spend a disproportionate amount of time and effort on a certain type of data (e.g., a system weakness).
By making users aware of the potential mistrust caused by providing feedback, they may be able to adjust their level of trust accordingly.

\subsection{Limitations and Future Work}
This research contributed empirical evidence that providing feedback can negatively effect user trust and perception of accuracy, but our findings also motivate the need to explore different kinds of feedback systems.
While our study focused on mandatory feedback to ensure that all participants engaged with the system equally, it is also possible to allow users to provide feedback optionally---an approach used in many intelligent systems.
In many such systems, users only provide feedback when they are already inclined to.
It is possible that removing the requirement to provide feedback could change the effects it has on user trust.
Therefore, more research is needed to fully understand and compare the impacts of mandatory and optional feedback on user trust and perception of accuracy.

Another potential direction for further research in this field would be to explore different methods of collecting human feedback.
Implicit feedback systems detect user behavior to update models without directly asking for input~\cite{shivaswamy2012online}.
However, with systems using this feedback technique, users might be aware that their behaviors are being recorded.
It may be interesting to see how this knowledge can affect trust and whether it is similar to our findings and observations.
Furthermore, it might be worthwhile to study whether this type of feedback can improve users' understanding of system accuracy and the impact of the their feedback usage for the model.
Additionally, while participants in our study provided feedback for individual system outputs, explanatory interactive learning systems can also allow users to modify model features directly~\cite{teso2019explanatory}.
The differences in these systems suggest that further research can study whether our findings extend to systems that use feature-based feedback.
Along these lines, future research may also consider whether the role of algorithmic transparency and system explanation might influence the user biases and perception of accuracy.
Thus, continued studies may also incorporate evaluation measures for explainable systems and understandability~\cite{mohseni2018survey} as an essential element of human-in-the-loop experiences.

\section{Conclusion}
\label{sec:conclusion}
Human-in-the-loop machine learning has many benefits, including the potential for increased model performance and providing users a way to control the outcomes of autonomous systems.
We conducted an experiment of the effects that providing such feedback has on users of intelligent systems in the presence of differing levels of change in accuracy over time.
The results show that regardless of the actual observed changes in system performance over time, participants who provided human-in-the-loop feedback believed the system to be significantly less accurate than participants who did not provide feedback.
The study also suggests participants who provided feedback trusted the system less than those who did not.
Therefore, developers of autonomous systems may need to consider the effects that allowing end users to provide feedback could have on how people perceive their models.

\section{Acknowledgements}
This work was supported by the DARPA Explainable Artificial Intelligence (XAI) Program under contract number N66001-17-2-4032 and by NSF award 1900767.

\bibliographystyle{aaai}
\bibliography{bibliography.bib}
\end{document}